\documentclass[aps,pra,superscriptaddress,twocolumn,amsmath,amssymb,showpacs,floatfix]{revtex4-1}

\usepackage{times}

\usepackage[utf8x]{inputenc}
\usepackage[english]{babel}
\usepackage[T1]{fontenc}
\usepackage{lmodern}
\usepackage{amssymb}
\usepackage{bbold}
\usepackage{amsmath}
\usepackage{amsthm}
\usepackage[pdftex]{graphicx}
\usepackage{epstopdf}

\usepackage{dcolumn}
\usepackage{bm}
\usepackage{color}
\usepackage{relsize,dsfont,mathrsfs,empheq,verbatim,upgreek}
\usepackage{etoolbox}
\usepackage[caption = false]{subfig}
\usepackage{pgfplots}

\newcommand{\bra}[1] {\left \langle #1 \right | }
\newcommand{\ket}[1] { \left | #1 \right \rangle }

\newcommand{\tonda}[1] { \left ( #1 \right ) }

\newcommand{\trn}[1] {  \vert \vert    #1 \vert \vert  }

\DeclareMathOperator{\Tr}{Tr}

\begin{document}


\title{Generalized trace distance approach to quantum non-Markovianity and detection of initial correlations}

\author{Giulio Amato}
\affiliation{Physikalisches Institut, Universit\"at Freiburg, 
Hermann-Herder-Stra{\ss}e 3, D-79104 Freiburg, Germany}
\affiliation{Dipartimento di Fisica, Universit\`a degli Studi di Milano,
Via Celoria 16, I-20133 Milan, Italy}

\author{Heinz-Peter Breuer}
\affiliation{Physikalisches Institut, Universit\"at Freiburg, 
Hermann-Herder-Stra{\ss}e 3, D-79104 Freiburg, Germany}

\author{Bassano Vacchini}
\affiliation{Dipartimento di Fisica, Universit\`a degli Studi di Milano,
Via Celoria 16, I-20133 Milan, Italy}
\affiliation{INFN, Sezione di Milano, Via Celoria 16, I-20133 Milan, Italy}

\date{\today}

\begin{abstract}
A measure of quantum non-Markovianity for an open system dynamics, based on revivals of the distinguishability between system states, has been introduced in the literature using the trace distance as quantifier for distinguishability.  Recently it has been suggested to use as measure for the distinguishability of quantum states the trace norm of Helstrom matrices, given by weighted differences of statistical operators. Here we show that this new approach, which generalizes the original one, is consistent with the interpretation of information flow between the system and its environment associated to the original definition.  To this aim we prove a bound on the growth of the external information, that is information which cannot be accessed by performing measurements on the system only, as quantified by means of the Helstrom matrix. We further demonstrate by means of example that it is of relevance in generalizing schemes for the local detection of initial correlations based on the increase of internal information. Finally we exploit this viewpoint to show the optimality of a previously introduced strategy for the local detection of quantum correlations.
\end{abstract}


\maketitle


\section{Introduction}

The standard description of an open quantum system dynamics rests on two basic assumptions: initial system-environment states in factorized form and weak coupling between the system and the environment \cite{Breuer2002}. The former request guarantees the existence of a reduced dynamics, while the latter introduces a separation of time scales between the evolution of the system and the environment so that a semigroup composition law in time can be reasonably adopted and the dynamics is fully characterized via its generator, given in Gorini-Kossakowski-Sudarshan-Lindblad form \cite{Gorini1976a,Lindblad1976a}. A lot of effort, in recent years, has been devoted to overcome these limitations.

Major results have been obtained in describing the dynamics outside the weak coupling regime, for initially factorized states, leading to reduced dynamics which go beyond the semigroup composition law \cite{BrLa2016}. In this regard different definitions of quantum non-Markovianity have been proposed \cite{BLP2009,RHP2010,Chru2011,WiVa2015}, with the goal of characterizing the set of quantum processes describing the time evolution of an open quantum system in terms of the produced memory effects. One promising and well established approach, also amenable to experimental testing \cite{Liu2011,Liu2013,Wittemer2017}, is the one based on the time evolution of distinguishability between pairs of open system states, to which is associated a meaning of information that can be extracted performing measures on the open system only. Quantum processes, obtained from global unitary evolutions tracing over the environment degrees of freedom, which lead to a monotonic decrease of such information regardless of the choice of initial reduced states are called Markovian, while non-Markovian are the processes which show a revival of information for at least one pair of initial system states. This definition can be connected to a property of Markovian classical stochastic processes, which lead to a monotonic decrease of Kolmogorov distance between probability vectors \cite{Vacchini2012a}.

However, the condition of initially factorized states between the system and the environment is rather limiting and a lot of efforts have been put forth to introduce reduced dynamical maps in the presence of initial correlations \cite{Pechukas1994,Rodriguez2008a,Shabani2009a,Brodutch2013a,Dominy2016b,VacAm2016}.
Nonetheless, such an extension is non trivial and severe limitations are encountered, so that a general and satisfactory treatment still lags behind. Conversely, general schemes for the detection of initial correlations through local measurements have been designed, using the aforementioned idea of flow of information \cite{LPB2010b,Gessner2011a,GBB2017,GBB2017}.

The distinguishability between quantum states thus plays a fundamental role both in theory of quantum non-Markovianity and in schemes for the local detection of initial correlations. The trace distance was initially adopted to quantify the distinguishability \cite{LPB2010a,Smirne2010,VaSm2011,Breuer2012a,BrAmVa2018}, relying on the paradigm of two states one-shot discrimination procedure presented in \cite{Nielsen2000,Fuchs1999}. However, more recently, in order to improve the definition of quantum non-Markovianity, it has been proposed to replace the trace distance between states with the trace norm of their weighted difference \cite{Chru2011}, the so-called Helstrom matrix, in line with \cite{Helstrom1976}: this choice enables a clear-cut connection to the classical definition and also a characterization of memory effects of a quantum process through its divisibility character \cite{WiVa2015}, which can be assessed looking the associated time-local generator.

The main aim of this paper is to show that the interpretation of flow of information between the system and the environment still holds adopting the trace norm of Helstrom matrices in place of the unbiased trace distance. This is obtained by introducing a new bound for the growth of external information as quantified via Helstrom matrices, generalizing the one obtained for the trace distance. This finding is fundamental to establish the generalized definition of quantum non-Markovianity, and also leads to the generalization of schemes for the local detection of initial system-environment correlations.



The paper is organized as follows: in Section \ref{sec:flow}, after a brief review of the theory of discrimination of quantum states, we will derive a general inequality for the so-called external information. This inequality proves to be very important to derive generalizations of schemes for the local detection of initial correlations to which Section \ref{sec:detection} will be devoted. In Section \ref{subsec:theory} we present the general theory, while in Section \ref{subsec:comparison} an example of spin-boson dynamics is analyzed, where a better detection capability is witnessed when using the novel approach, which enables to detect initial correlations in the global system-environment states when the trace distance approach fails. Furthermore in Section \ref{subsec:qcor} a generalization of a method to detect quantum correlations is also presented, whose optimality can be assessed in this new framework, as shown with an explicit example in Section \ref{subsec:twoqubit}. 


\section{Generalized trace distance approach to information flow and quantum non-Markovianity}
\label{sec:flow}
Given a quantum system with an associated Hilbert space $ \mathcal{H} $, we describe its states by density matrices $ \rho $, i.e. positive trace class operators with unit trace, whose set is denoted by $ \mathcal{S}(\mathcal{H}) $. Henceforth, we denote by $ \trn{A} = \Tr | A |$ the trace norm of $ A$, where the modulus of an operator is defined via $ | A | = \sqrt{A^{\dagger} A} $.

\subsection{Distinguishability between states and information in a quantum system}
\begin{figure}
\centering
\includegraphics[width=0.5\textwidth]{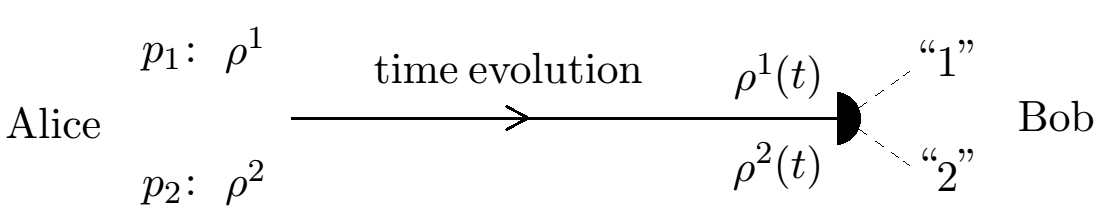}
\caption{Cartoon of the two states one-shot discrimination procedure: Alice prepares a quantum system in the states $ \rho^1 ( \rho^2) $ with probabilities $ p_1 (p_2 ) $, whereas $ ''{1}'' $ and $ ''{2}'' $ are the possible results of Bob's measurement. Increasing the number of possible results of the measurement does not lead to an improved distinguishability of the prepared states.}
\label{fig:discrim}
\end{figure}

To introduce the concept of information, which proves to be fundamental in all this work, we consider two parties, Alice and Bob \cite{BLP2009,WiVa2015}: Alice prepares the system in two possible quantum states $ \rho^1 $ and $ \rho^2 $, with respective probabilities $ p_1 $ and $ p_2 $, with $ p_1 , p_2 \geq 0 $ and $ p_1 + p_2 = 1 $, and then sends it to Bob. Bob has the task to figure out, by means of a single measurement, 
whether the system has been prepared in the state $ \rho^1 $ or $ \rho^2 $ (see Figure \ref{fig:discrim}). The success probability in the discrimination procedure, if the measurement is carried out by means of two positive operators $ \{ P_1, P_2 \} $ such that $ P_1 , P_2 \geq 0 $ and $ P_1 + P_2 = I $, is 
\begin{equation*}
p_{\textrm{success}} (t) =  p_1 \Tr [ P_1 \rho^1 (t) ] + p_2  \Tr [ P_2 \rho^2 (t) ] .
\label{eq:pSucc}
\end{equation*}
Expressing $ P_1 $ in function of $ P_2 $ and vice versa one gets
\begin{equation}
p_{\textrm{success}} (t) = p_2 + \Tr [ P_1 \Delta (t) ] = p_1 - \Tr [ P_2 \Delta (t) ] ,
\label{eq:pSucc1}
\end{equation}
where we have defined the Helstrom matrix
\begin{equation}
\Delta (t) = p_1 \rho^1 (t) - p_2 \rho^2 (t) . 
\label{eq:helstrom}
\end{equation}
Hence, one can recast \eqref{eq:pSucc1} as
\begin{equation*}
\begin{aligned}
p_{\textrm{success}} (t) = & \; \frac{1}{2} \mathlarger{[} p_2 + \Tr [ P_1 \Delta (t) ] + p_1 - \Tr [ P_2 \Delta (t) ]  \mathlarger{]} \\
= & \; \frac{1}{2} \mathlarger{[} 1 + \Tr [ ( P_1 - P_2 ) \Delta (t) ]  \mathlarger{]} ,
\end{aligned}
\label{eq:pSucc3}
\end{equation*}
which is maximal if $ P_1 $ and $ P_2 $ are the projectors on the subspaces spanned by the positive and negative eigenvectors of $ \Delta (t) $ respectively, and, in that case, is equal to
\begin{equation*}
p^{\textrm{max}}_{\textrm{success}} (t) = \frac{1}{2} \mathlarger{[} 1 + \trn{ \Delta (t) } \mathlarger{]}.
\label{eq:pSuccMax}
\end{equation*}
The trace norm of the Helstrom matrix is connected to the information which can be obtained measuring the evolved states, because it provides the bias in favor of their distinguishability. Note that considering initial equal preparation frequencies $ p_1 = p_2 = 1/2 $, the trace norm of \eqref{eq:helstrom} reduces to the trace distance between the states \cite{Nielsen2000}
\begin{equation*}
D ( \rho^1 (t) , \rho^2 (t) ) = \frac{1}{2} \Tr | \rho^1 (t) - \rho^2 (t) | ,
\label{eq:trDist}
\end{equation*}
which is maximal and equal to one for orthogonal states, i.e. with orthogonal support, while zero for equal density matrices.

The evolution in time of the distinguishability between the initial states prepared by Alice is related to the character of the dynamics. If one considers a closed and isolated quantum system, such as the total system $ S+E $, whose evolution is described by time-dependent unitary operators $ \{ U_t \}_t $, acting on density operators as $ \rho_{SE} (t) = U_t \rho_{SE} U_t^{ \dagger} $, $ \trn{ p_1 \rho_{SE}^1 (t) - p_2 \rho_{SE}^2 (t) } = \trn{ p_1 \rho_{SE}^1 - p_2 \rho_{SE}^2 } $. In fact unitary transformations do not change the spectrum of operators. We define this conserved quantity as total information in the composite system $ S + E $
\begin{equation*}
 \mathcal{I }_{\textrm{tot}} (t) = \trn{ p_1 \rho_{SE}^1 (t) - p_2 \rho_{SE}^2 (t) } .
\label{eq:totInfo}
\end{equation*}
However, the open quantum system $ S $ undergoes non-unitary dynamics, derived from the underlying global coherent evolution by tracing over the environmental degrees of freedom
\begin{equation}
\rho_S = \Tr_E \rho_{SE} \textrm{ } \mapsto \textrm{ } \rho_S (t) = \Tr_E [ U_t \rho_{SE} U_t^{ \dagger} ] .
\label{eq:sEvo}
\end{equation} 
Hence, the distinguishability between reduced states can vary in time. We define the internal information as the information accessible by Bob, if he is allowed to perform measurements on the open quantum system $S $ only
\begin{equation*}
 \mathcal{I }_{\textrm{int}} (t) = \trn{ p_1 \rho_{S}^1 (t) - p_2 \rho_{S}^2 (t) } .
\label{eq:intInfo}
\end{equation*}
Consequently, we define the external information as
\begin{equation*}
\begin{aligned}
\mathcal{I }_{\textrm{ext}} (t) = & \textrm{ } \mathcal{I }_{\textrm{tot}} (t) - \mathcal{I }_{\textrm{int}} (t) \\
= & \textrm{ } \trn{ p_1 \rho_{SE}^1 (t) - p_2 \rho_{SE}^2 (t) } - \trn{ p_1 \rho_{S}^1 (t) - p_2 \rho_{S}^2 (t) } .
\end{aligned}
\label{eq:extInfo}
\end{equation*}

Hence, the conservation of the total information, in terms of internal and external information at different times $ t ,s \geq 0 $, can be written as a balance equation
\begin{equation}
\mathcal{I }_{\textrm{int}} (t) + \mathcal{I }_{\textrm{ext}} (t) = \mathcal{I }_{\textrm{int}} (s) + \mathcal{I }_{\textrm{ext}} (s) .
\label{eq:consTot}
\end{equation}

\subsection{Inequality for the external information}
In this section we present an inequality for the external information, as captured by the trace norm of Helstrom matrices, which generalizes the inequality derived for the trace distance case \cite{BrLa2016}. This is the main finding of this article. The inequality represents an important result because it allows to extend the interpretation of non-Markovianity in terms of the flow of information between the open system and its environment to case when the distinguishability is measured in terms of the trace norm of the Helstrom matrix.

The external information at a given time $ t $ is nonzero if correlations in the global states $ \rho_{SE}^{1,2} (t) $ are present or the environmental marginals are different, as expressed by the following inequality
\begin{equation}
\begin{aligned}
\mathcal{I }_{\textrm{ext}} (t) \leq & \; 2 p_1 D ( \rho_{SE}^1 (t) , \rho_{S}^1 (t) \otimes \rho_{E}^1 (t)  ) \\
+ & \; 2 p_2 D ( \rho_{SE}^2 (t) , \rho_{S}^2 (t) \otimes \rho_{E}^2 (t)) \\
+ & \; 2 \min \{ p_1, p_2 \} D ( \rho_{E}^1 (t) , \rho_{E}^2 (t) ) .
\end{aligned}
\label{eq:extInfoIneq}
\end{equation}
The trace distance case \cite{BrLa2016} can be trivially retrieved by setting $ p_1 = p_2 = 1/2 $. The proof is given in the appendix.

\subsection{Generalized trace distance approach to quantum non-Markovianity}

The possibility to introduce reduced dynamical maps, i.e. maps acting on the open quantum system degrees of freedom only describing its dynamics, treating effectively the presence of the environment, rests on the condition of initially factorized states $ \rho_{SE} = \rho_{S} \otimes \rho_E $, with a fixed marginal state of the environment. In fact, in such case the quantum process $ \mathlarger{ \Phi} $ is given by a collection of trace-preserving time-dependent maps $ \{ \Phi_t \}_t $, which are completely positive \cite{Stinespring1955,Choi1972}, defined via
\begin{equation*}
\Phi_t \rho_S =  \Tr_E [ U_t \rho_S \otimes \rho_E U_t^{\dagger} ] .
\label{eq:redDynamics} 
\end{equation*}
\textbf{Definition.} A quantum process $ \mathlarger{ \Phi} $ is Markovian if $ \trn{ p_1 \Phi_t \rho_S^1 - p_2 \Phi_t \rho_S^2 } $ is a monotonic decreasing function of time, for any $ p_1 , p_2 \geq 0 $ with $ p_1 + p_2 = 1 $ and $ \rho_S^{1}, \rho_S^{2} \in \mathcal{ S} ( \mathcal{ H}_S ) $ \cite{WiVa2015}.\\
$ \textrm{ } $\\ 
An interpretation via information flow between the system and the environment has been devised for the original definition of quantum non-Markovianity \cite{BLP2009,LPB2010a}, and as shown here via Eq. \eqref{eq:extInfoIneq} it can be generalized for the new definition in a straightforward way.

In case of initially factorized states $ \rho_{SE}^{1,2} = \rho_S^{1,2} \otimes \rho_E $, which is a premise to introduce reduced dynamical maps and, hence, talk about a quantum process, we have that total and internal information at initial time do coincide. In fact, we have 
\begin{equation*}
\begin{aligned}
 \mathcal{I }_{\textrm{tot}} (0) = & \textrm{ } \trn{ p_1 \rho_{S}^1 \otimes \rho_E - p_2 \rho_{S}^2  \otimes \rho_E }\\
= & \textrm{ } \trn{ p_1 \rho_{S}^1 - p_2 \rho_{S}^2 } \textrm{ } \trn{ \rho_E}  =  \mathcal{I }_{\textrm{int}} (0)
\end{aligned}
\label{eq:intTotEqual}
\end{equation*} 
and, consequently, $ \mathcal{I }_{\textrm{ext}} (0) = 0 $. Thus, recasting the equation for the conservation of the total information \eqref{eq:consTot}, choosing $ s = 0 $ the initial time, we have
\begin{equation*}
\mathcal{I }_{\textrm{int}} (0) - \mathcal{I }_{\textrm{int}} (t) = \mathcal{I }_{\textrm{ext}} (t) \geq 0 ,
\label{eq:noIncremInt}
\end{equation*}
which express the fact that the internal information is always bounded from above by its initial value. This is of course in agreement with the contraction property of the trace norm under (completely) positive and trace preserving reduced dynamical maps \cite{Ruskai1994}.

Let us now consider the equation for the conservation of the total information \eqref{eq:consTot}: a backflow of information, captured by an increase of the internal information between $ s > 0 $ and the later time $ t $, can be present only if the external information is non zero at the former time. In fact
\begin{equation}
\begin{aligned}
\mathcal{I }_{\textrm{int}} (t) - \mathcal{I }_{\textrm{int}} (s) = & \textrm{ } \mathcal{I }_{\textrm{ext}} (s) - \mathcal{I }_{\textrm{ext}} (t) \\
\leq & \textrm{ } \mathcal{I }_{\textrm{ext}} (s) ,
\end{aligned}
\label{eq:recastConsTot}
\end{equation}
because $ \mathcal{I }_{\textrm{ext}} (t) $ is a positive quantity. The interpretation of the external information, for this generalized definition with the trace norm of Helstrom matrices, as presence of correlations in the states $ \rho_{SE}^{1,2} (s) $ or as different environmental states is expressed by inequality \eqref{eq:extInfoIneq}.


\section{Generalization of schemes for the local detection of initial correlations}
\label{sec:detection}
\subsection{Theoretical analysis}
\label{subsec:theory}
In case the assumption of initially factorized states is violated, there is in general no possibility to introduce reduced dynamical maps and hence speak of a quantum process, let alone of Markovianity and non-Markovianity. However, the interpretation via flow of information is still valid and can be used to witness locally the presence of initial correlations in the global system-environment states \cite{LPB2010b}. Here we show a generalization of the bound presented in \cite{BrLa2016,LPB2010b}, considering the trace norm of Helstrom matrices in place of the trace distances of states. Setting $ s = 0 $ on right-hand side of \eqref{eq:recastConsTot}, we get
\begin{equation}
\begin{aligned}
\mathcal{I }_{\textrm{int}} (t) - \mathcal{I }_{\textrm{int}} (0) = & \textrm{ } \mathcal{I }_{\textrm{ext}} (0) - \mathcal{I }_{\textrm{ext}} (t) \\
\leq & \textrm{ } \mathcal{I }_{\textrm{ext}} (0)
\end{aligned}
\label{eq:intInfoIncrease}  
\end{equation}
and, by means of \eqref{eq:extInfoIneq}, we obtain the generalized bound for the increase of internal information with respect to its initial value
\begin{equation}
\begin{aligned}
\mathcal{I }_{\textrm{int}} (t) - \mathcal{I }_{\textrm{int}} (0) 
\leq & \textrm{ } 2 p_1 D ( \rho_{SE}^1 , \rho_{S}^1  \otimes \rho_{E}^1  ) \\
+ & \textrm{ } 2 p_2 D ( \rho_{SE}^2 , \rho_{S}^2 \otimes \rho_{E}^2 ) \\
+ & \textrm{ } 2 \min \{ p_1, p_2 \} D ( \rho_{E}^1 , \rho_{E}^2 ) ,
\end{aligned}
\label{eq:intInfoBound}
\end{equation}
which links a possible increase of internal information above the intial value to different environmental initial states or correlations in the total initial states.

\subsection{Performance comparison of different approaches: an example}
\label{subsec:comparison}
To show that the generalized bound (3.2) can indeed lead to a better sensitivity in detecting initial correlations we provide an explicit example.

We consider a two-level system subjected to a pure dephasing dynamics, generated by the interaction with a bosonic mode, through the full Hamiltonian
\begin{equation}
H = H_S \otimes I_E + I_S \otimes H_E + S \otimes X,
\label{eq:puredephasingham}
\end{equation}
where $ S \otimes X $ is the interaction term between the system and the environment satisfying
\begin{equation*}
[ H_S , S ] = 0.
\label{eq:puredephasingcondition}
\end{equation*}
This condition ensures that $ H_S $ is a conserved quantity, because $ [ H_S, H ] = 0 $ and enables to carry out analytical calculations. In particular, we take in (\ref{eq:puredephasingham})
\begin{equation*}
\begin{aligned}
H_S = & \textrm{ } \epsilon \sigma_S^3 , \\
H_E = & \textrm{ }  \omega b^{\dagger} b ,\\
S \otimes X =  & \textrm{ }  g \sigma_S^3 \otimes ( b + b^{\dagger}),
\end{aligned}
\label{eq:myChoiceH}
\end{equation*}
where $ b $ and $ b^{\dagger} $ are lowering and raising operators of the bosonic mode, $ g $ is the coupling constant, while $ \epsilon $ and $ \omega $ are the energy spacing between the eigenstates of the system and the environment respectively, in the units where $ \hbar = 1 $. We consider, as initial condition for our dynamical evolution, states depending on the parameter $ \lambda \in [0,1] $ of the form \cite{Dajka2011,WiLe2013}
\begin{equation}
\ket{ \Psi_{\lambda} (0) }_{SE} = \alpha \ket{ 1} \otimes \ket{0}_E + \beta \ket{0 } \otimes \ket{ \Omega_{ \lambda} }_E , 
\label{eq:dephinitialstate}
\end{equation}
where $ \ket{ k } $, with $ k = 1,0$, are respectively the excited and ground states of the two level system, while $ \ket{0}_E $ and $\ket{ \Omega_{ \lambda} }_E $ are states of the bosonic mode. We require $ \vert \alpha \vert^2 + \vert \beta \vert^2 = 1 $ to ensure normalization of $ \ket{ \Psi_{\lambda} (0) }_{SE}$. The field state is given by the coherent superposition of the vacuum state of the bosonic mode $ \ket{0}_E $ and a certain coherent state $ \ket{y}_E $, that is
\begin{equation}
\ket{ \Omega_{ \lambda} }_E = \frac{1}{C_{\lambda} } [ ( 1 - \lambda) \ket{0}_E + \lambda  \ket{y}_E  ],
\label{eq:statoomegadeph}
\end{equation}
with $ b \ket{y}_E = y \ket{y}_E $, while the normalization factor reads
\begin{equation*}
C_{\lambda} = \sqrt{ ( 1 - \lambda)^2 + \lambda^2 + 2 \lambda (1 - \lambda) \textrm{ Re}( { }_E \langle 0 \vert y \rangle_E ) } .
\label{eq:normFactor}
\end{equation*}
This state reduces to the vacuum state for $ \lambda = 0 $, so that the initial composite state (\ref{eq:dephinitialstate}) is factorized. Otherwise, aside from $ \alpha = 0 $ or $ \beta = 0 $, we have an initial state $ \ket{ \Psi_{\lambda} (0) }_{SE} $ that is not factorized, but is actually entangled.

The dynamical evolution of the reduced state of the system \eqref{eq:sEvo} can be studied through the evolution of its matrix elements $ \bra{i } \rho_S (t) \ket{j } $, where $ \ket{i}, \ket{j} $ are the eigenvectors of the system free Hamiltonian $ H_S = \epsilon \sigma_S^3 $. The expression has been derived in \cite{Dajka2011,WiLe2013} and, upon choosing the coherent state $ \ket{ y}_E $ in $ \ket{ \Omega_{\lambda} }_E $ (\ref{eq:statoomegadeph}) to be $  \ket{ y = 1}_E $, reads
\begin{equation*}
\rho^{\lambda}_S (t) = \begin{pmatrix}
\vert \alpha \vert^2 & \alpha \beta^* B_{\lambda} (t) \\
 \alpha^* \beta B^*_{\lambda} (t) & \vert \beta \vert^2 
\end{pmatrix} ,
\label{eq:dynevodeph}
\end{equation*}
where $ \rho_S^{\lambda} (t) = \Tr_E [ \ket{ \Psi_{ \lambda} (t) }_{SE} \bra{ \Psi_{ \lambda} (t) } ] $, with
\begin{equation*}
B_{ \lambda } (t) = \frac{1}{C_{ \lambda} } e^{-2 i \epsilon t }  e^{ - R (t) } \mathlarger [ 1- \lambda + \lambda e^{ - 2 i \Lambda (t) + S (t) }  \mathlarger ] , 
\label{eq:parametro}
\end{equation*}
and
\begin{equation}
\begin{aligned}
R (t) = & \textrm{ } 4 \tonda{ \frac{g}{ \omega} }^2 [ 1 - \cos ( \omega t) ] , \\
\Lambda (t) = & \textrm{ } \frac{g}{ \omega} \sin (\omega t ) , \\
S (t) = & \textrm{ } 2 \frac{g}{ \omega} [ 1 - \cos ( \omega t) ] - \frac{1}{2} .
\end{aligned}
\label{eq:parametrioscillanti}
\end{equation}

Therefore, to study the efficiency of this local scheme for the detection of initial correlations, we consider the reduced system states $ \rho_S^0 $, which is initially factorized, and $ \rho_S^{\lambda } $, with the parameter $ \lambda $ entailing the quantity of correlations in the initial global state. Their Helstrom matrices
\begin{widetext}
\begin{equation}
\begin{aligned}
\Delta (t) = & \textrm{ } p_1 \rho_S^{\lambda} (t) - p_2 \rho_S^0 (t) \\
= & \textrm{ } \begin{pmatrix}
( p_1 - p_2 ) \vert \alpha \vert^2 & \alpha \beta^* e^{-2 i \epsilon t } e^{ - R (t) }  \mathlarger [ p_1 C_{\lambda}^{-1} \mathlarger ( 1 - \lambda + \lambda e^{ - 2 i \Lambda (t) + S (t) } \mathlarger ) - p_2  \mathlarger ]  \\
h.c.  & ( p_1 - p_2 ) \vert \beta \vert^2
\end{pmatrix} ,
\end{aligned}
\label{eq:trnhexw}
\end{equation}
\end{widetext}
are functions of $ p_1 $, $ \alpha $ and $ \beta $, upon choosing the physical parameters of the model as $ \epsilon = 1 $, $ \omega = 1 $ and $ g = 0.1 $, in agreement with the literature \cite{Dajka2011,WiLe2013}.

We carry out numerical simulations to find the probability of detecting initial correlations with the novel approach, comparing the results with the performance of the old approach used in \cite{WiLe2013}. It is important to note that, in such article, the authors confront the efficiency of different distance measures, namely the trace distance, the Bures metric, the Hellinger distance and the Jensen-Shannon divergence \cite{Hayashi2006}, with the first one being the most effective: here thus we compare the trace distance with the trace norm of the Helstrom matrix.

\begin{figure}[tbh]
\centering
\includegraphics[width=0.5\textwidth]{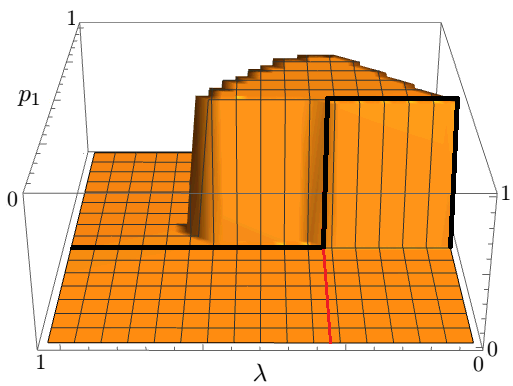}
\caption{Probability of detection of initial correlations for different values of $ \{ \alpha , \beta \} $, for $ 40 $ and $ 30 $ equally spaced values of $ p_1 $ and $ \lambda $ respectively. The black line is the result shown in \cite{WiLe2013} (Figure 2 therein), where the authors considered the trace distance to detect initial correlations, i.e. $ p_1 = 0.5 $. We see that, after the threshold value $ \lambda \gtrsim 0.4 $, highlighted by a red line, initial correlations are detected only choosing $ p_1 > 1/2 $ and hence using the novel approach.}
\label{fig:dal}
\end{figure}

To this end, for each considered value of $ p_1 \in [ 0,1]  $ and $ \lambda \in [ 0,1]  $, we determine the probability of an increase in time of the internal information with respect to its initial value: we  randomly draw $ 500 $ pairs $ \{\alpha, \beta \} $ in their range of variability and determine the frequency of detection of initial correlations. The maximum time considered in this computation is $ t = 2 \pi $, corresponding to the period of the oscillating quantities \eqref{eq:parametrioscillanti} in \eqref{eq:trnhexw}, and the evolved internal information is computed with a time step $ \textrm{d}t = 0.15 $. The result is depicted in Figure \ref{fig:dal}.
The graph shows a threshold value in detection of initial correlations, when using the trace distance \cite{WiLe2013}: in fact, for $ \lambda \gtrsim 0.4 $ and $ p_1 = 1/2 $ there is no increase of internal information with respect to its initial value. Nonetheless, if one considers the novel approach and chooses $ p_1 > 1/2 $ - giving, hence, \textit{more weight} to the correlated state in the Helstrom matrix \eqref{eq:trnhexw} - one can better detect initial correlations, witnessing them for $ \lambda > 0.4 $.

\begin{figure}[tbh]
\centering
\includegraphics[width=88mm]{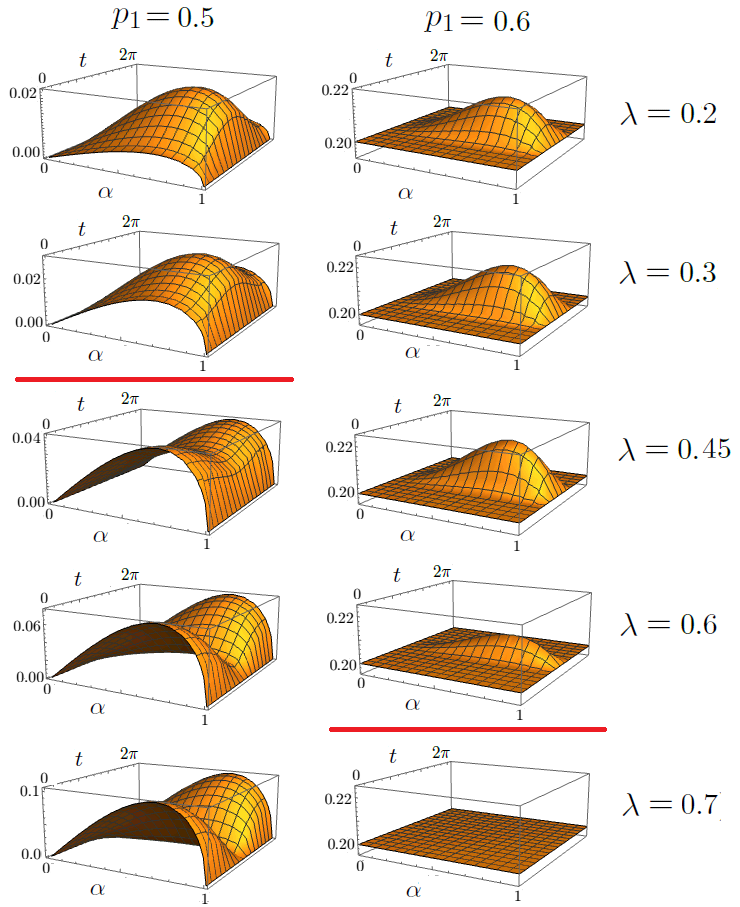}
\caption{Behaviour of the trace norm of (\ref{eq:trnhexw}) for $ p_1 = 0.5 $ and $ p_1= 0.6 $, for different values of $ \lambda $, as functions of $ \alpha $ and $ t $. On the left side, related to the trace distance case $ p_1 = 0.5 $, no increase of distinguishability is detected for $ \lambda \gtrsim 0.4 $, while the trace norm with $ p_1 = 0.6 $ shows an increase also for larger values of $ \lambda $. The red line separates a positive witness of initial correlations from a failure of the method, for $ p_1 =0.5 $ and $ p_1 =0.6 $.}
\label{fig:graficone}
\end{figure}

To stress this better detection capability, we plot, for different values of $ \lambda $ around the threshold value $ 0.4 $, the trace distance and the trace norm of (\ref{eq:trnhexw}) with $ p_1 = 0.6 $ as functions of time and the free parameters $ \{ \alpha, \beta \} $. In this case $ \alpha, \beta \in \mathbb{R} $, with $ \beta = \sqrt{ 1 - | \alpha \vert^2 } $. As we can see from Figure \ref{fig:graficone}, no initial increase is observed after the threshold value $ \lambda \sim 0.4$ using the trace distance, while the novel approach with $ p_1 = 0.6 $ stops witnessing correlations around $ \lambda \sim 0.7 $.

\subsection{Local detection method for quantum correlations}
\label{subsec:qcor}
The idea of information has been interestingly used also to devise a method to detect quantum correlations \cite{qd0,qd1,qd2} in a composite system using only local operations \cite{Gessner2011a,Gessner2013a}. This method has been experimentally realized for both trapped ions \cite{Gessner2014a} and photonic systems \cite{Cialdi2014a,Tang2015}.

The basic idea behind such method is to consider the evolution of the internal information between the composite state of interest $ \rho_{SE} $ and the state $ \rho'_{SE} = ( \Phi_d \otimes I_E ) \rho_{SE} $, where $ \Phi_d $ is the local map acting as
\begin{equation*}
\Phi_d X = \sum_{j=1}^{n } \Pi_S^j X \Pi_S^j ,
\label{eq:depGess}
\end{equation*}
with $ \{ \Pi_S^j \}_{j=1,...,n } $ the complete set of rank-one orthogonal projectors on the eigenstates of $ \rho_S $, namely $ \rho_S = \sum_{j=1}^{n} q_j \Pi_S^j $, where $ n = \textrm{dim} \mathcal{H}_S $. Thus the corresponding composite state change is given by
\begin{equation}
\rho'_{SE} = \sum_{j=1}^n q_j \Pi_S^j \otimes \rho_E^j ,
\label{eq:0QDstateGess}
\end{equation}
with
\begin{equation}
\rho_E^j = \frac{ \Tr_S [ ( \Pi_S^j \otimes I_E ) \rho_{SE} ( \Pi_S^j \otimes I_E ) ]  }{ \Tr_{SE} [ ( \Pi_S^j \otimes I_E ) \rho_{SE}  ]  } 
\label{eq:helpRicRho}
\end{equation}
and
\begin{equation}
q_j = \Tr_{SE} [ ( \Pi_S^j \otimes I_E ) \rho_{SE} ] .
\label{eq:helpRicQ}
\end{equation}
Note that $ \Phi_d $ can be built through local operations, i.e. only performing measurements on the reduced state.

Although the open system states are identical at the initial time, their global counterparts $ \rho_{SE} $ and $\rho'_{SE} $ are identical only if $ \rho_{SE} $ is originally a zero quantum discord state of the form \eqref{eq:0QDstateGess}. This leads to possible differences in the time evolution of the reduced open system states, so that
\begin{equation}
D ( \rho_S (t), \rho'_S (t) ) = \frac{1}{2} \trn{ \Tr_E [ U_t ( \rho_{SE} - \rho'_{SE} ) U_t^{\dagger} ] } > 0 
\label{eq:witnessGessner}
\end{equation}
has been considered as a witness for quantum correlations in $ \rho_{SE} $.

Here, we demonstrate that the local detection method can be generalized by using the norm of Helstrom matrices as quantifiers for state discrimination. 
Thus, we consider $ \mathcal{I }_{\textrm{int}} (t) = \trn{ p_1 \rho_S (t) - p_2 \rho'_S (t) } $, and we propose as a novel condition for the detection of initial quantum correlations
\begin{equation}
\mathcal{I }_{\textrm{int}} (t) - \mathcal{I }_{\textrm{int}} (0) > 0 ,
\label{eq:newMethod}
\end{equation}
where, in particular, we have $ \mathcal{I }_{\textrm{int}} (0) = \vert p_1 - p_2 \vert $. 

The interpretation via flow of information between the system and the environment is captured by the following upper bound for the increase of internal information proven in appendix
\begin{equation}
\mathcal{I }_{\textrm{int}} (t) - \mathcal{I }_{\textrm{int}} (0) \leq 2 \min \{ p_1, p_2 \} \textrm{ }  D ( \rho_{SE} , \rho'_{SE} ) ,
\label{eq:newMethodBound}
\end{equation}
so that the maximum increase of internal information is captured via the trace distance between $ \rho_{SE} $ and the classically correlated $ \rho_{SE}'$.

It is important to notice that this novel approach leads to no improvements in the detection of quantum correlations. In fact, on the one hand, both witnesses \eqref{eq:witnessGessner} and \eqref{eq:newMethod} are able to detect initial correlations if and only if 
\begin{equation*}
\rho_S (t) \neq \rho_S'(t) 
\label{eq:witCond}
\end{equation*}
for some $ t > 0 $. Thus, the use of the trace norm of the Helstrom matrix does not improve the capability of detecting quantum correlations. On the other hand, the bound for the increase of internal information, given by inequality \eqref{eq:newMethodBound}, is maximal for $ p_1 = p_2 = 1/2 $, which indicates that the new witness is not more sensitive than the witness based on the trace distance, which therefore turns out to be the optimal one.

\subsection{A two qubit example}
\label{subsec:twoqubit}
We now focus on a specific example with low dimensional system and environment. This enables the evaluation of the r.h.s. of the bound \eqref{eq:newMethodBound}, which requires the computation of the trace distance of system and environment states. We thus consider the model originally proposed in \cite{LPB2010b} consisting of a couple of two-level systems undergoing a CNOT quantum gate \cite{Nielsen2000}. Thus we will not study the continuum time evolution of the internal information, but we will just compute and compare the initial value $ \mathcal{I}_{\textrm{int}}(0) $, with the one after the state transformation has occurred, that we still indicate with $ \mathcal{I}_{\textrm{int}}(t) $. The action of the CNOT can be defined by
\begin{equation*}
\begin{aligned}
U_C  \ket{ 1 1 } = & \ket{ 1 1 }, \quad \quad   U_C  \ket{ 1 0 } =  \ket{ 1 0 } ,\\
U_C  \ket{ 0  1 } = & \ket{ 0 0 } , \quad \quad  U_C  \ket{ 0 0  } =  \ket{ 0 1 } ,
\end{aligned}
\label{eq:truthCNOT}
\end{equation*}
where $ \ket{ij} \bra{kl} = \ket{i} \bra{k} \otimes \ket{j} \bra{l} $ for any $i,j,k,l \in \{0, 1\} $ and $ \sigma^3 \ket{j} = (-1)^{j+1} \ket{j} $. We consider the families of initial states \cite{StBuzek2001}
\begin{subequations}
\begin{align}
\rho_{SE} = & \textrm{ } ( \alpha \ket{ 1 1  } + \beta \ket{0 0 } ) ( \alpha^* \bra{ 1 1 } + \beta^* \bra{ 0 0} ) \label{eq:statesStBu1} \\
\rho'_{SE} = & \textrm{ } \vert \alpha \vert^2 \ket{ 11  } \bra{ 11} + \vert \beta \vert^2 \ket{00 } \bra{ 00} \label{eq:statesStBu2},  
\end{align}
\end{subequations}
with $ \vert \alpha \vert^2 + \vert \beta \vert^2 +  = 1 $, which are respectively pure entangled states and their classically correlated version, i.e. $ \rho'_{SE} = \Phi_d \rho_{SE} $, as explained in Sec. \ref{subsec:qcor} (see Appendix).

\begin{figure}
\centering
 \includegraphics[width=65mm]{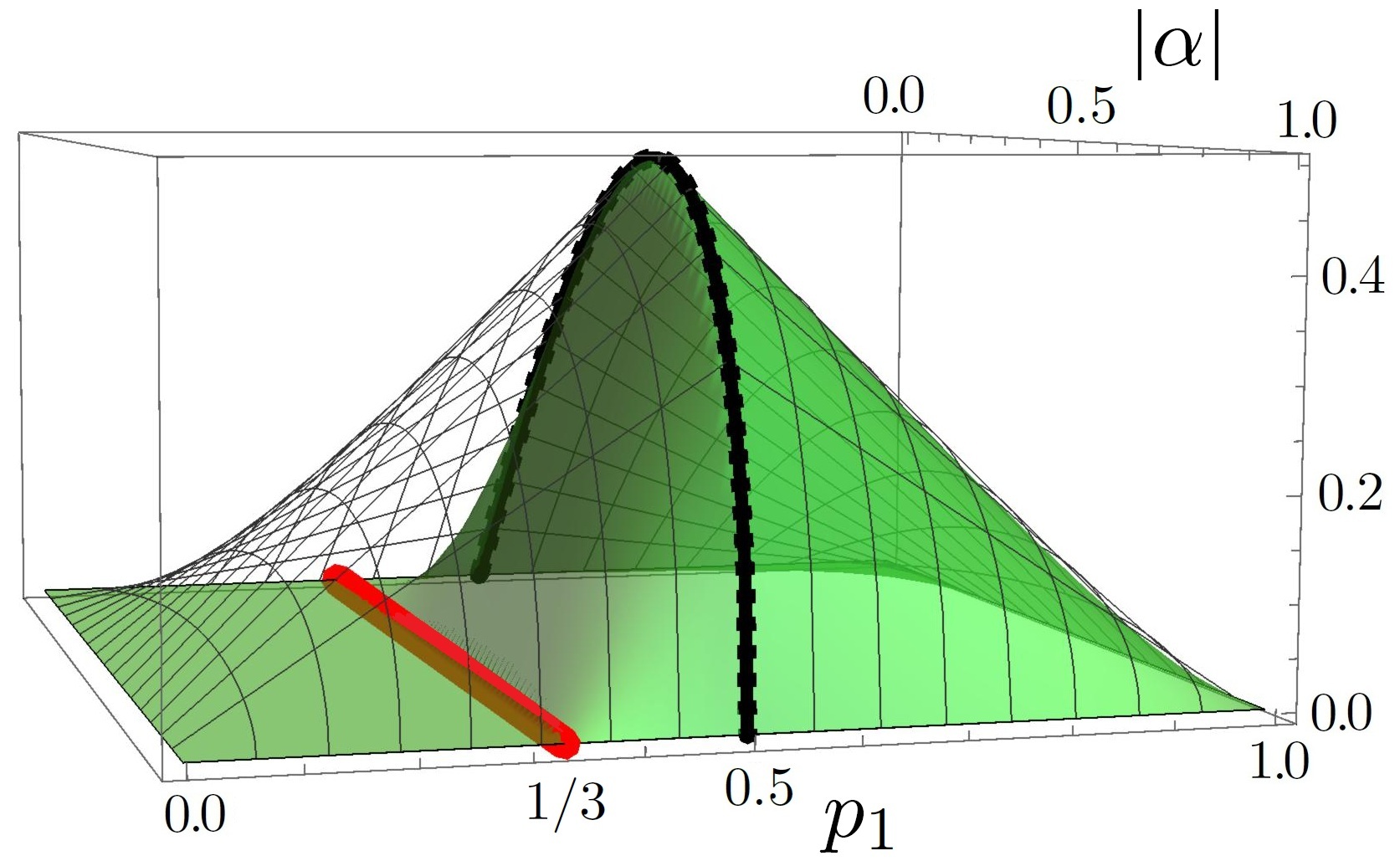}
\caption{In green (gray) we plot the increase of internal information $ \mathcal{I}_{\textrm{int}} (t) - \mathcal{I}_{\textrm{int}} (0) $, while the meshed surface is the bound \eqref{eq:valBoundEasy}, as a function of $ \vert \alpha \vert \in [0,1]$ and $ p_1 \in [0,1] $. Note that the maximum increase is given by the unbiased case $ p_1 = 1/2 $, that is highlighted by a black line, where the bound is also saturated: in this situation, the information contained in the quantum correlations is completely transferred, via the CNOT, into the system. No initial quantum correlations are detected for $p_1 < 1/3 $, as indicated by the red line.}
 \label{fig:b2}
\end{figure}

We know that $ \mathcal{I}_{\textrm{int}} (0) =  \vert p_1 - p_2 \vert $. The calculation of the evolved internal information is quite lengthy and worked out in the appendix with the result
\begin{equation}
\begin{aligned}
& \mathcal{I}_{\textrm{int}} (t) \\
\textrm{ } & = \begin{cases} 
   \vert p_1 - p_2 \vert  & \text{if } p_1 < 1/3  \\
   \sqrt{ ( p_1 - p_2)^2 ( \vert \alpha \vert^2 - \vert \beta \vert^2 )^2 + 4 p_1^2 \vert \alpha \beta \vert^2  }  & \text{if } p_1 \geq 1/3 .
  \end{cases}
\end{aligned}
\label{eq:unicaimportante}
\end{equation}
The bound \eqref{eq:newMethodBound}, also computed in the appendix, reads
\begin{equation}
2 \min \{ p_1, p_2 \} D ( \rho_{SE} , \rho'_{SE} ) = 2 \min \{ p_1, p_2 \} \textrm{ }  \vert \alpha \beta \vert .
\label{eq:valBoundEasy}
\end{equation} 
We represent the increase of internal information and its bound in Figure \ref{fig:b2}. The plot is in accordance with the theoretical predictions, presented in the previous section: there are no initial quantum correlations that can be detected via the method which adopts the trace norm of Helstrom matrices, which cannot be detected using the trace distance. Moreover, the maximum increase of internal information is obtained, for any choice of the parameters $\alpha, \beta $ characterizing the state \eqref{eq:statesStBu1}, when $ p_1 = p_2 = 1/2$. In particular, in this example, there is no detection of initial quantum correlations when $p_1 < 1/3 $, as indicated by the red line.

\section{Conclusions}
\label{sec:conclusion}
In this paper we have highlighted the importance of the notion of information flow as a unifying paradigm to treat non-Markovianity of quantum processes and to devise schemes for the local detection of initial correlations in open quantum systems, showing in particular that this concept is robust with respect to different formalizations.

The information flow between a quantum system and other degrees of freedom it is interacting with is studied by means of the Helstrom matrix, thus following a recent generalization of the trace distance criterion, as proposed in \cite{Chru2011,WiVa2015}.  The Helstrom matrix is given by the weighted difference of two statistical operators, so that its norm coincides with the trace distance between the operators if the weights are equal. We show in particular that a fundamental inequality, used to provide a bound on the information which due to the interaction between system and environment is no more available performing measurements on the system only, can be naturally extended from the trace distance approach to its generalization based on the Helstrom matrix. This simple result allows to prove that all results and interpretations used within the trace distance approach can be consistently applied to its generalization.

We show in particular by means of an explicit example of a spin-boson model, that the trace norm of Helstrom matrices is more effective in witnessing the presence of initial correlations by means of local information with respect to the trace distance.

At the same time we prove that a recently introduced scheme for the local detection of quantum correlations based on the trace distance approach \cite{Gessner2011a,GBB2017} is indeed optimal, in the sense that the generalized witness obtained in the Helstrom matrix formulation does detect the very same correlations and exhibit its greatest sensitivity for the unbiased case in which the two statistical operators are considered with equal weight.

\section*{ACKNOWLEDGMENTS}
HPB and BV acknowledge support from the European Union (EU) through the Collaborative Project QuProCS (Grant Agreement No. 641277). BV further acknowledges support from FFABR.

\section*{APPENDIX}

\subsection{Proof of Eq. \eqref{eq:extInfoIneq}}
Using the property $ \trn{ A \otimes B } = \trn{A } \textrm{ } \trn{B }$ and because $ \trn{ \rho } = 1 $ for any statistical operator, we have
\begin{equation}
\begin{aligned}
\mathcal{I}_{\textrm{ext}} (t) = & \textrm{ } \trn{ p_1 \rho_{SE}^1 (t) - p_2 \rho_{SE}^2 (t) } \\
& \textrm{ } - \trn{ ( p_1 \rho_{S}^1 (t) - p_2 \rho_{S}^2 (t) ) \otimes \rho_E^1 (t) } ,
\end{aligned}
\label{eq:1case}
\end{equation}
but also
\begin{equation}
\begin{aligned}
\mathcal{I}_{\textrm{ext}} (t) = & \textrm{ } \trn{ p_1 \rho_{SE}^1 (t)  - p_2 \rho_{SE}^2 (t) } \\
 & \textrm{ } - \trn{ ( p_1 \rho_{S}^1 (t) - p_2 \rho_{S}^2 (t) ) \otimes \rho_E^2 (t) } .
\end{aligned}
\label{eq:2case}
\end{equation}
Exploiting positivity of $ \mathcal{I}_{\textrm{ext}} (t) $ from \eqref{eq:1case}, using the triangular inequality twice, we obtain
\begin{equation*}
\begin{aligned}
& \mathcal{I}_{\textrm{ext}} (t) \\ 
& \textrm{ } \leq \trn{ p_1 \rho_{SE}^1 (t) - p_2 \rho_{SE}^2 (t) - ( p_1 \rho_{S}^1 (t) - p_2 \rho_{S}^2 (t) ) \otimes \rho_E^1 (t) } \\
& \textrm{ } =  \trn{ p_1 [ \rho_{SE}^1 (t) - \rho_{S}^1 (t) \otimes \rho_{E}^1 (t) ] - p_2 [ \rho_{SE}^2 (t) - \rho_{S}^2 (t) \otimes \rho_E^1 (t) ] } \\
& \textrm{ }  \leq 2 p_1 D ( \rho_{SE}^1 (t), \rho_{S}^1 (t) \otimes \rho_{E}^1 (t) ) + 2 p_2 D ( \rho_{SE}^2 (t), \rho_{S}^2 (t) \otimes \rho_E^1 (t) ) .
\end{aligned}
\label{eq:1caseCont}
\end{equation*}
Using the triangular inequality for the trace distance and the property
\begin{equation*}
D ( \rho_S^2 (t) \otimes \rho_E^2 (t) , \rho_{S}^2 (t) \otimes \rho_E^1 (t) ) = D ( \rho_E^2 (t) ,\rho_E^1 (t) ) ,
\label{eq:trdHelp1}
\end{equation*}
we, hence, obtain
\begin{equation*}
\begin{aligned}
\mathcal{I}_{\textrm{ext}} (t) \leq & \textrm{ } 2 p_1 D ( \rho_{SE}^1 (t), \rho_{S}^1 (t) \otimes \rho_{E}^1 (t) ) \\
+ & \textrm{ } 2 p_2 D ( \rho_{SE}^2 (t), \rho_S^2 (t) \otimes \rho_E^2 (t) ) \\
+ & \textrm{ }  2 p_2 D ( \rho_E^1 (t) ,\rho_E^2 (t) ) .
\end{aligned} 
\label{eq:1caseContFine}
\end{equation*}
Analogously, from \eqref{eq:2case}, we obtain
\begin{equation*}
\begin{aligned}
\mathcal{I}_{\textrm{ext}} (t) \leq & \textrm{ } 2 p_1 D ( \rho_{SE}^1 (t), \rho_{S}^1 (t) \otimes \rho_{E}^1 (t) ) \\
+ & \textrm{ } 2 p_2 D ( \rho_{SE}^2 (t), \rho_S^2 (t) \otimes \rho_E^2 (t) ) \\
+ & \textrm{ } 2 p_1 D ( \rho_E^1 (t) ,\rho_E^2 (t) ) ,
\end{aligned} 
\label{eq:2caseContFine}
\end{equation*}
and therefore the claimed result \eqref{eq:extInfoIneq}.

\subsection*{Proof of Eq. \eqref{eq:newMethodBound}}
We consider inequality \eqref{eq:intInfoIncrease}, where in this case
\begin{equation*}
\mathcal{I }_{\textrm{ext}} (0) = \trn{ p_1 \rho_{SE} - p_2 \rho'_{SE} } - | p_1 - p_2 | ,
\label{eq:infExtCase}
\end{equation*}
because $ \rho_S = \rho_S' $ and $ \mathcal{I }_{\textrm{int}} (0) $ reduces to $ | p_1 - p_2 | $. In particular, because $ | x | = | x | \textrm{ } \trn{ \rho } $ for any density operator $ \rho $ and $ x \in \mathbb{ C} $, we have
\begin{equation*}
\mathcal{I }_{\textrm{ext}} (0) = \trn{ p_1 \rho_{SE} - p_2 \rho'_{SE} } - \trn{ (p_1 - p_2 ) \rho_{SE} } ,
\label{eq:infExtCaseH1}
\end{equation*}
but also
\begin{equation*}
\mathcal{I }_{\textrm{ext}} (0) = \trn{ p_1 \rho_{SE} - p_2 \rho'_{SE} } - \trn{ (p_1 - p_2 ) \rho'_{SE} } .
\label{eq:infExtCaseH2}
\end{equation*}
In both cases, because $ \mathcal{I }_{\textrm{tot}} (0) \geq \mathcal{I }_{\textrm{int}} (0) $, we can use the triangular inequality for the trace norm in the form
\begin{equation*}
\big{|} \textrm{ } \trn{A } - \trn{ B} \textrm{ } \big{|} \leq \trn{ A - B } ,
\label{eq:revTriangular}
\end{equation*}
which leads, via \eqref{eq:intInfoIncrease}, to
\begin{equation*}
\mathcal{I }_{\textrm{int}} (t) -  \mathcal{I }_{\textrm{int}} (0) \leq \trn{ p_2 ( \rho_{SE} - \rho'_{SE} ) }
\label{eq:prefinDim1}
\end{equation*}
and
\begin{equation*}
\mathcal{I }_{\textrm{int}} (t) -  \mathcal{I }_{\textrm{int}} (0) \leq \trn{ p_1 ( \rho_{SE} - \rho'_{SE} ) }
\label{eq:prefinDim2}
\end{equation*}
Hence, by definition of trace distance, we obtain the general bound \eqref{eq:newMethodBound}.

\subsection*{Derivation of Eqs. \eqref{eq:unicaimportante} and \eqref{eq:valBoundEasy}}
Here we show the calculations needed to derive the results \eqref{eq:unicaimportante} and \eqref{eq:valBoundEasy}.

To this end, we first check the relation $ \rho'_{SE} = (\Phi_d \otimes I_E ) \rho_{SE} $ for the global states defined in \eqref{eq:statesStBu1} and \eqref{eq:statesStBu2}. Because the marginal state is $ \rho_S = \Tr_E \rho_{SE} = \vert \alpha \vert^2 \ket{0} \bra{0 } + \vert \beta \vert^2 \ket{1} \bra{1 } $, we have
\begin{equation*}
\Phi_d X = \sum_{j=1,0} \Pi_S^j X \Pi_S^j ,
\label{eq:phidex}
\end{equation*}
with $ \Pi_S^1 = \ket{1} \bra{1} $ and $ \Pi_S^0 = \ket{0} \bra{0} $. Computing $ q_j $ as in \eqref{eq:helpRicQ}, we have
\begin{equation*}
q_1 = \vert \alpha \vert^2 , \quad q_0 = \vert \beta \vert^2 ,
\end{equation*}
while $ \rho_E^j $ as in \eqref{eq:helpRicRho} are
\begin{equation*}
\begin{aligned}
\rho_E^1 = & \textrm{ } \frac{1}{ \vert \alpha \vert^2 } \vert \alpha \vert^2 \ket{1} \bra{1} = \ket{1} \bra{1} , \\
\rho_E^0 = & \textrm{ } \frac{1}{ \vert \beta \vert^2 } \vert \beta \vert^2 \ket{0} \bra{0} = \ket{0} \bra{0} .
\end{aligned}
\end{equation*}
It is easy to see that $ \rho'_{SE} = (\Phi_d \otimes I_E ) \rho_{SE} $.\\

Secondly we compute the result of Eq. \eqref{eq:unicaimportante}. The transformation of the states $ \rho_{SE} $, as in \eqref{eq:statesStBu1}, and $ \rho'_{SE} $, as in \eqref{eq:statesStBu2}, under the CNOT is
\begin{equation*}
\begin{aligned}
U_C \rho_{SE} U_C^{\dagger} = & \textrm{ } ( \alpha \ket{ 11 } + \beta \ket{01} ) \textrm{ }  ( \alpha^* \bra{ 11 } + \beta^* \bra{0 1} ) , \\
U_C \rho'_{SE} U_C^{\dagger} = & \textrm{ }  \vert \alpha \vert^2 \ket{ 11  }  \bra{ 11 } + \vert \beta \vert^2 \ket{0 1} \bra{0 1} .
\end{aligned}
\label{eq:stBuzTrans}
\end{equation*} 
Hence, the transformed marginals of the system are
\begin{equation*}
\begin{aligned}
\Tr_E [ U_C \rho_{SE} U_C^{\dagger} ] = & \textrm{ }  \vert \alpha \vert^2 \ket{ 1 } \bra{ 1} + \vert \beta \vert^2 \ket{0} \bra{ 0} \\
+ & \textrm{ } \alpha \beta^*  \ket{ 1} \bra{ 0} + \alpha^* \beta \ket{0} \bra{ 1} ,  \\
\Tr_E [ U_C \rho'_{SE} U_C^{\dagger}] = & \textrm{ } \vert \alpha \vert^2 \ket{ 1 } \bra{ 1} + \vert \beta \vert^2 \ket{0} \bra{ 0}  .
\end{aligned}
\end{equation*}
Thus, we have that $ \mathcal{I }_{\textrm{int}} (t) = \trn{ p_1 \Tr_E [ U_C \rho_{SE} U_C^{\dagger} ] - p_2 \Tr_E [ U_C \rho'_{SE} U_C^{\dagger} ] } $ is equal to the trace norm of the Helstrom matrix
\begin{equation}
\Delta = \begin{pmatrix}
(p_1 - p_2) \vert \alpha \vert^2 & p_1 \alpha \beta^* \\
p_1 \alpha^* \beta & (p_1 - p_2) \vert \beta \vert^2
\end{pmatrix} = \begin{pmatrix}
\gamma & \theta \\
\theta^* & \delta
\end{pmatrix} ,
\label{eq:trnCalc}
\end{equation}
having introduced
\begin{equation*}
\begin{aligned}
\gamma = & \textrm{ } (p_1 - p_2) \vert \alpha \vert^2  ,\\
\delta = & \textrm{ } (p_1 - p_2) \vert \beta \vert^2 , \\
\theta = & \textrm{ } p_1 \alpha \beta^* .
\end{aligned}
\label{eq:trnCalcHelp}
\end{equation*}
The eigenvalues of \eqref{eq:trnCalc} are
\begin{equation*}
\eta_{\pm} = \frac{ \gamma + \delta \pm \sqrt{ ( \gamma - \delta )^2 + 4 \vert \theta \vert^2 } }{2} ,
\label{eq:trnCalcEig}
\end{equation*}
so that the trace norm of \eqref{eq:trnCalc} is
\begin{equation}
\begin{aligned}
& \trn{ \Delta }  \\
& =  \begin{cases} 
   \vert \gamma + \delta \vert  & \quad \text{if } \vert  \gamma + \delta \vert  > \sqrt{ ( \gamma - \delta )^2 + 4 \vert \theta \vert^2 } \\
   \sqrt{ ( \gamma - \delta )^2 + 4 \vert \theta \vert^2 }       & \quad  \text{if } \vert  \gamma + \delta \vert \leq \sqrt{ ( \gamma - \delta )^2 + 4 \vert \theta \vert^2 }
  \end{cases}.
\end{aligned}
\label{eq:trnCalcPartial}
\end{equation}
To express \eqref{eq:trnCalcPartial} in terms of the original parameters $ \alpha $, $ \beta $ and $ p_1 $, we note that $ \gamma + \delta = p_1 - p_2 $ and
\begin{equation*}
( \gamma - \delta )^2 + 4 \vert \theta \vert^2 = ( p_1 - p_2)^2 ( \vert \alpha \vert^2 - \vert \beta \vert^2 )^2 + 4 p_1^2 \vert \alpha \beta \vert^2 ,
\label{eq:trnCalcHelpIndietro}
\end{equation*}
so that the separating condition in \eqref{eq:trnCalcPartial} reads 
\begin{equation}
 ( p_1 - p_2)^2 -( p_1 - p_2)^2 ( \vert \alpha \vert^2 - \vert \beta \vert^2 )^2 + 4 p_1^2 \vert \alpha \beta \vert^2 \lessgtr 0 .
\label{eq:condValHestrom}
\end{equation}
Because $ \vert \alpha \vert^2 + \vert \beta \vert^2 = 1 $, we can choose the parametrization $ \vert \alpha \vert = \cos (z) $ and $ \vert \beta \vert = \sin (z)  $, with $ z \in [0,\pi/2] $, so that (\ref{eq:condValHestrom}) corresponds to
\begin{equation*}
\begin{aligned}
0 & \lessgtr \textrm{ } ( p_1 - p_2)^2 - ( p_1 - p_2)^2 \cos^2 (2z) - p_1^2 \sin^2 (2z) \\
& = \textrm{ } \sin^2 (2z) [ ( p_1 - p_2)^2 - p_1^2 ] .
\end{aligned}
\end{equation*}
Thus, trivially, one obtains $ ( 2 p_1 - 1 )^2 - p_1^2 \lessgtr 0 $, which leads to the desired result \eqref{eq:unicaimportante}.\\

Finally, to obtain the value of the bound \eqref{eq:valBoundEasy} in the considered example, we compute the trace distance between $ \rho_{SE} $ and $ \rho'_{SE} $, defined respectively in \eqref{eq:statesStBu1} and \eqref{eq:statesStBu2}, which corresponds to the sum of the moduli of the eigenvalues of $  1/2 [ \rho_{SE} - \rho'_{SE} ] $. The only non zero entries of this $ 4 \times 4 $ matrix are the coherences of $ \rho_{SE} $ and it is easy to see that $ D (\rho_{SE}, \rho'_{SE}) = \vert \alpha \beta \vert $.

\end{document}